\documentclass[letterpaper]{article} 
\usepackage{aaai2026}  
\usepackage{times}  
\usepackage{helvet}  
\usepackage{courier}  
\usepackage[hyphens]{url}  
\usepackage{graphicx} 
\usepackage{natbib}  
\usepackage{caption} 
\usepackage{algorithm}
\usepackage{algorithmic}
\usepackage{amsmath}
\usepackage{amsfonts}
\usepackage{booktabs}
\usepackage{enumitem}
\usepackage{float}
\usepackage{newfloat}
\usepackage{listings}
\DeclareCaptionStyle{ruled}{labelfont=normalfont,labelsep=colon,strut=off} 
\floatstyle{ruled}
\newfloat{listing}{tb}{lst}{}
\floatname{listing}{Listing}
\title{DeepBayesFlow: A Bayesian Structured Variational Framework for Generalizable Prostate Segmentation via Expressive Posteriors and SDE-Girsanov Uncertainty Modeling}
\author {
    Zhuoyi Fang
}
\affiliations{
    School of Mathematical Sciences\\
    Yangzhou University\\
    230802107@stu.yzu.edu.cn
}

\usepackage{bibentry}

\begin{document}

\maketitle

\begin{abstract}
Automatic prostate MRI segmentation faces persistent challenges due to inter-patient anatomical variability, blurred tissue boundaries, and distribution shifts arising from diverse imaging protocols. To address these issues, we propose DeepBayesFlow, a novel Bayesian segmentation framework designed to enhance both robustness and generalization across clinical domains. DeepBayesFlow introduces three key innovations: a learnable NF-Posterior module based on normalizing flows that models complex, data-adaptive latent distributions; a NCVI inference mechanism that removes conjugacy constraints to enable flexible posterior learning in high-dimensional settings; and a SDE-Girsanov module that refines latent representations via time-continuous diffusion and formal measure transformation, injecting temporal coherence and physically grounded uncertainty into the inference process. Together, these components allow DeepBayesFlow to capture domain-invariant structural priors while dynamically adapting to domain-specific variations, achieving accurate and interpretable segmentation across heterogeneous prostate MRI datasets.

\end{abstract}

\section{Introduction}

The prostate, as a vital male reproductive organ, plays an essential role in urinary and sexual functions. Prostate cancer remains one of the most common malignancies worldwide, underscoring the critical importance of accurate prostate MRI segmentation for early diagnosis, treatment planning, and disease monitoring \cite{azad2024medical,he2023accuracy,muller2022towards}. The segmentation task is particularly challenging due to inherent anatomical variability among patients, where prostate size and shape can differ widely. Additionally, the internal zonal anatomy presents subtle contrast differences \cite{ramesh2021review,rayed2024deep,chen2024transunet}, and lesions often appear with blurred boundaries and heterogeneous textures, complicating precise delineation \cite{valanarasu2022unext,zhang2023customized}.

Deep learning-based segmentation methods, especially those derived from the U-Net family, have revolutionized prostate MRI analysis by leveraging hierarchical feature extraction and skip connections that preserve spatial resolution \cite{butoi2023universeg,xing2024segmamba,isensee2024nnu}. Subsequent innovations have integrated attention mechanisms and Transformer architectures to capture both local and global contextual information \cite{wu2025medical,chen2024ma,hatamizadeh2022unetr}, thereby enhancing the detection of fine-grained anatomical structures and improving boundary delineation in prostate imaging \cite{zhang2024segment,roy2023mednext,wu2024medsegdiff}. These advances have pushed the performance limits on curated datasets with relatively consistent acquisition protocols \cite{salpea2022medical,xiao2023transformers}.

Despite significant progress in prostate MRI segmentation, existing methods often struggle to generalize robustly across diverse clinical scenarios characterized by wide anatomical variability, ambiguous tissue boundaries, and imaging artifacts. The prostate exhibits substantial inter-patient variation in shape and size, with lesion-prone zones frequently presenting diffuse edges and heterogeneous textures. Moreover, subtle intensity differences between healthy and pathological tissues pose additional challenges for conventional feature-based segmentation approaches. Real-world MRI data further complicate this task due to noise, motion-induced artifacts, and partial volume effects, necessitating segmentation frameworks that are both uncertainty-aware and capable of adapting to distributional shifts. 

To address the challenges of generalization and uncertainty in prostate MRI segmentation, we propose DeepBayesFlow—a principled Bayesian framework designed to enhance both predictive performance and interpretability. DeepBayesFlow introduces three key innovations: NF-Posterior, which employs normalizing flows to model rich, input-dependent posterior distributions that go beyond the limitations of traditional Gaussian assumptions; NCVI, a flexible inference strategy that removes conjugacy constraints, allowing for more accurate posterior estimation across diverse anatomical structures; and the SDE-Girsanov latent diffusion mechanism, which refines latent representations over time through stochastic differential equations and measure transformation, ensuring temporal coherence and physical plausibility. Together, these components form an integrated Bayesian pipeline capable of capturing complex, multimodal uncertainties while preserving high segmentation fidelity across variable imaging conditions and pathological diversity.

DeepBayesFlow is built upon an encoder-decoder architecture with skip connections and a lightweight probabilistic head, enabling precise boundary detection and well-calibrated uncertainty estimation. This design supports robust generalization across domains that differ in acquisition quality, protocols, and patient demographics. By modeling uncertainty throughout the segmentation process, DeepBayesFlow delivers reliable, interpretable predictions, paving the way for trustworthy deployment in real-world clinical settings.

Our main contributions include:
\begin{enumerate}[label=\arabic*)]
\item The development of \textit{DeepBayesFlow}, a novel Bayesian segmentation model that unifies flow-based posterior learning, non-conjugate variational inference, and stochastic latent diffusion to enhance generalization in prostate MRI segmentation.
\item The introduction of \textit{NF-Posterior}, a flexible posterior approximation module based on normalizing flows, capable of capturing complex anatomical and pathological uncertainty beyond conventional Gaussian assumptions.
\item The design of \textit{NCVI}, a non-conjugate inference framework that decouples prior and posterior forms, enabling robust performance under anatomical variability, noise, and domain shifts.
\item The formulation of \textit{SDE-Girsanov}, a stochastic differential equation-driven latent diffusion process that injects temporal coherence and physically grounded uncertainty refinement across MRI slices.
\item Comprehensive evaluations on multiple public prostate MRI datasets demonstrating that DeepBayesFlow consistently achieves state-of-the-art performance and superior generalization to unseen patient cohorts and imaging conditions.
\end{enumerate}

\section{Related work}

\subsection{Deep learning for prostate image segmentation}
\label{subsec:segmentation}

Deep learning has significantly advanced semantic segmentation for prostate imaging, with methods grouped into four categories: feature encoder-based, upsampling-based, feature resolution enhancement, and region proposal-based techniques \cite{liu2021review,orlando2022ultrasound,wang2023dsunet}.

These methods extract hierarchical features using convolutional networks like ResNet and Xception, which address deep network training challenges via residual connections and efficient convolutions \cite{conze2023trends}. Recent models such as EfficientNet and transformer-based architectures improve representation by balancing model complexity and capturing global context \cite{wang2023twostage}. To recover spatial details lost during downsampling, approaches like FCN, U-Net, and SegNet employ upsampling and skip connections that combine low- and high-level features. Advances include learnable upsampling and attention mechanisms to enhance spatial precision \cite{chen2021alexnet}. Atrous (dilated) convolution expands the receptive field without reducing resolution, enabling multi-scale context capture as seen in DeepLab variants \cite{jia2022synergic}. DeepLabV3+ further refines segmentation by fusing features at different scales, integrating efficient convolutions to reduce computational cost \cite{bhandary2023unet}. Region proposal networks like Faster R-CNN and Mask R-CNN generate candidate object regions and perform pixel-level segmentation with precise spatial alignment \cite{jiang2023ultrasound}. These frameworks have been adapted for medical imaging to accurately detect and segment anatomical regions and lesions \cite{sammouda2021kmeans,zaridis2023region}.

These evolving techniques underpin current state-of-the-art prostate segmentation models, balancing accuracy, efficiency, and robustness.

\subsection{Domain generalization for medical image segmentation}
\label{subsec:generalization}

Domain generalization (DG) focuses on training models using one or multiple source domains that can generalize well to unseen target domains without any access to target data during training \cite{yoon2024domain}. This setting is more challenging than domain adaptation but better reflects real-world clinical applications where target domain data is unavailable \cite{ouyang2022causality}.

DG methods are generally categorized into data-based, learning-based, and representation-based approaches \cite{hu2022domain}. Data-based methods enhance model robustness by increasing training data diversity through augmentations or synthetic generation, simulating domain shifts, imaging artifacts, or signal corruptions \cite{su2023rethinking}. Techniques include stacked transformations, adversarial augmentations, mutual information regularization, and causality-inspired data augmentation, all aimed at encouraging domain-invariant feature learning \cite{liu2021feddg}.

Learning-based strategies improve the training process itself, often through meta-learning frameworks designed to promote shape compactness, boundary awareness, and disentanglement of domain-specific factors \cite{xu2022adversarial}. Episodic training schemes in frequency or other transformed domains fuse information from multiple sources to build more generalizable models \cite{yao2022enhancing}. These approaches encourage the model to adapt dynamically to unseen domain variations \cite{gao2024desam}.

Representation-based approaches aim to extract domain-invariant features by minimizing distribution discrepancies across domains via adversarial learning, statistical metrics such as Maximum Mean Discrepancy, or entropy regularization \cite{lyu2022aadg}. Feature disentanglement further separates domain-specific and domain-agnostic components \cite{li2022domain}. Recent progress also integrates causal inference to identify latent factors truly influencing predictions, enhancing robustness to domain shifts \cite{wang2022rethinking}. Among such methods, BayeSeg \cite{gao2023bayeseg} stands out by modeling domain invariance within a Bayesian framework, offering interpretable and principled uncertainty estimation. While representation-based DG has made significant strides in general computer vision, its application to medical image segmentation remains limited, with interpretability of invariant features posing ongoing challenges for clinical adoption \cite{zhao2022efficient,hu2022domain,ouyang2022causality,liu2020shapeaware,zhang2019when}.

\begin{figure*}[htbp]
\centering
\includegraphics[width=\textwidth]{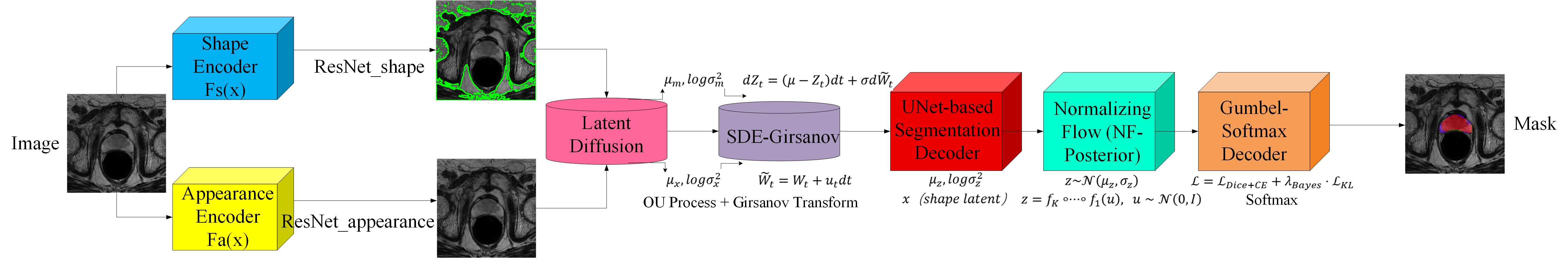}
\caption{Model Architecture Overview of DeepBayesFlow.}
\label{fig:tu1}
\end{figure*}

\section{Method}

\subsection{Model Overall}
\label{subsec:overall}

DeepBayesFlow is a principled Bayesian segmentation framework designed to address uncertainty, anatomical ambiguity, and domain variability in prostate MRI analysis. It features dual-stream encoders that disentangle structural and appearance priors and a segmentation decoder that integrates multi-source features for voxel-level predictions, enabling unified modeling of semantic consistency and spatial uncertainty. Key innovations include the NF-Posterior module, which uses normalizing flows to learn expressive, invertible latent transformations beyond Gaussian assumptions; Non-Conjugate Variational Inference (NCVI) that removes restrictive conjugacy constraints for flexible posterior adaptation; and the SDE-Girsanov mechanism, which employs stochastic differential equations to model latent variable evolution with learned drift corrections, providing a rigorous and interpretable uncertainty propagation. Together, these components form a mathematically sound and robust probabilistic inference engine, demonstrating superior performance in capturing fine anatomical details, quantifying predictive confidence, and generalizing across diverse imaging domains, making DeepBayesFlow a powerful tool for prostate MRI segmentation in both research and clinical applications.

\begin{algorithm}
\caption{DeepBayesFlow Inference with SDE-Girsanov, NF-Posterior, and NCVI}
\label{alg:deepbayesflow}
\begin{algorithmic}[1]

\STATE // Input: Image $I \in \mathbb{R}^{B \times 1 \times H \times W}$, Num classes $K$

\STATE // \textbf{1. Appearance Encoding and Latent Sampling via SDE-Girsanov}
\STATE $(\mu_m, \log \sigma^2_m) \leftarrow \text{ResNet\_appearance}(I)$
\STATE $\sigma_m \leftarrow \exp(0.5 \cdot \log \sigma^2_m)$
\STATE $m \leftarrow \text{SDE\_Girsanov}(\mu_m, \sigma_m)$

\STATE // \textbf{2. Shape Encoding and Latent Sampling via SDE-Girsanov}
\STATE $(\mu_x, \log \sigma^2_x) \leftarrow \text{ResNet\_shape}(I)$
\STATE $\sigma_x \leftarrow \exp(0.5 \cdot \log \sigma^2_x)$
\STATE $x \leftarrow \text{SDE\_Girsanov}(\mu_x, \sigma_x)$

\STATE // \textbf{3. Observation Likelihood (NCVI)}
\STATE $r \leftarrow I - (x + m)$
\STATE $\mu_{\rho} \leftarrow \frac{2\gamma_\rho + 1}{r^2 + 2\phi_\rho}$
\STATE Sample $n \leftarrow \text{SDE\_Girsanov}(m, \log(1/\mu_\rho))$

\STATE // \textbf{4. Segmentation Latent Inference}
\STATE $x' \leftarrow \text{RepeatChannels}(x, 3)$
\STATE $(\mu_z, \log \sigma^2_z) \leftarrow \text{EfficientUNet}(x')$
\STATE $\sigma_z \leftarrow \exp(0.5 \cdot \log \sigma^2_z)$
\STATE $z \leftarrow \text{SDE\_Girsanov}(\mu_z, \sigma_z)$

\STATE // \textbf{5. Normalizing Flow Posterior Refinement}
\IF{training}
  \STATE $z_{\text{flat}} \leftarrow \text{Reshape}(z, [-1, K])$
  \STATE $z_{\text{nf}} \leftarrow \text{NF\_Posterior}(z_{\text{flat}})$
  \STATE $z_{\text{nf}} \leftarrow \text{Reshape}(z_{\text{nf}}, [B, K, H, W])$
  \STATE $\hat{Y} \leftarrow \text{GumbelSoftmax}(z_{\text{nf}})$
\ELSE
  \STATE $\hat{Y} \leftarrow \text{Softmax}(z)$
\ENDIF

\STATE // \textbf{6. NCVI: Compute Variational Regularization Terms}
\STATE Compute $\mu_\upsilon \leftarrow \frac{2\gamma_\upsilon + K}{\sum \mu_z \cdot (||\nabla \mu_x||^2 + 2\sigma_x^2) + 2\phi_\upsilon}$
\STATE Compute $\mu_\omega \leftarrow \frac{2\gamma_\omega + 1}{\pi \cdot (||\nabla \mu_z||^2 + 2\sigma_z^2) + 2\phi_\omega}$
\STATE Compute $\pi \leftarrow \text{MeanSpatial}(\mu_z)$
\STATE $\alpha_\pi, \beta_\pi \leftarrow \text{UpdatedBetaPrior}(\mu_\omega, \nabla \mu_z, \sigma_z)$
\STATE $\psi \leftarrow \text{Digamma}(\alpha_\pi + \beta_\pi) - \text{Digamma}(\beta_\pi)$

\STATE // \textbf{7. KL Divergences for Loss}
\STATE $\text{KL}_y \leftarrow \sum \mu_\rho \cdot r^2$
\STATE $\text{KL}_z \leftarrow \sum \psi \cdot \mu_\omega \cdot ||\nabla \mu_z||^2 + 2\sigma_z^2$
\STATE $\text{KL}_x \leftarrow \sum \mu_z \cdot \mu_\upsilon \cdot ||\nabla \mu_x||^2 + 2\sigma_x^2$
\STATE $\text{KL}_m \leftarrow \sigma_0 \cdot (\mu_m^2 + \sigma_m^2)$

\STATE // \textbf{8. Return Prediction and Regularizers}
\STATE \textbf{return} $\hat{Y}, \text{KL}_y, \text{KL}_z, \text{KL}_x, \text{KL}_m, \mu_\rho, \mu_\omega, \mu_\upsilon$
\end{algorithmic}
\end{algorithm}

\subsection{Normalizing Flow for Posterior Modeling of Latent Variable $z$}
\label{subsec:normalizing-flow}

NF-Posterior is designed to overcome the restrictive Gaussian assumptions commonly imposed in standard variational frameworks, which are inadequate for modeling the rich posterior complexity present in prostate MRI segmentation. Such complexity arises from anatomical variations across patients, uncertain lesion boundaries, and modality-induced artifacts. To this end, NF-Posterior constructs an expressive approximate posterior by transforming a simple base distribution $u \sim \mathcal{N}(0, \mathbf{I})$ via a composition of $K$ invertible and differentiable mappings $f_k: \mathbb{R}^D \rightarrow \mathbb{R}^D$, yielding:
\begin{equation}
z = f_K \circ f_{K-1} \circ \cdots \circ f_1(u),
\end{equation}
where the resulting density under $q(z|x)$ is evaluated using the change-of-variable formula:
\begin{equation}
\log q(z|x) = \log q(u) - \sum_{k=1}^K \log \left| \det \left( \frac{\partial f_k}{\partial f_{k-1}} \right) \right|.
\end{equation}
Each $f_k$ is instantiated using Masked Autoregressive Transforms (MAF) combined with Reverse Permutation layers to ensure both high expressiveness and efficient Jacobian computation. During training, samples of $z$ are drawn from this flow-based posterior for reconstruction and segmentation learning; during inference, its mean $\mu_z$ is used for deterministic prediction. By explicitly modeling complex, input-conditioned uncertainty structures, NF-Posterior enhances the model’s capability to disambiguate fuzzy anatomical structures, such as prostate boundaries near the rectum or bladder wall.

\subsection{Non-Conjugate Variational Inference for Flexible Posterior Learning}
\label{subsec:nonconjugate-vi}

NCVI serves as the optimization framework underlying NF-Posterior, removing the classical constraint of conjugacy between the latent prior $p(z)$ and approximate posterior $q_\phi(z|x)$. While the prior remains a standard normal $p(z) = \mathcal{N}(0, \mathbf{I})$, the posterior is modeled by NF-Posterior, enabling a highly non-Gaussian, learnable structure. Training is conducted by maximizing the Evidence Lower Bound (ELBO):
\begin{equation}
\mathcal{L}_{\text{ELBO}} = \mathbb{E}_{q_\phi(z|x)}[\log p(y|z)] - \mathrm{KL}[q_\phi(z|x) \| p(z)],
\end{equation}
where the KL term is tractably computed via:
\begin{equation}
\mathrm{KL}[q_\phi(z|x) \| p(z)] = \mathbb{E}_{u \sim \mathcal{N}(0,\mathbf{I})} \left[ \log \frac{q_\phi(z|x)}{p(z)} \right],
\end{equation}
with $q_\phi(z|x)$ derived through the accumulated log-determinants of the normalizing flow. This unconstrained variational setting allows the posterior to flexibly match the complex latent structure induced by ambiguous prostate boundaries, heterogeneous intensities, or pathology-induced deviations. NCVI enhances the segmentation model’s adaptability to out-of-distribution cases, such as atypical gland shapes or rare lesion morphologies, by avoiding rigid prior-posterior coupling.

\subsection{Advanced SDE-based Latent Diffusion with Girsanov Transformation}
\label{subsec:advanced-sde}

SDE-Girsanov introduces a continuous-time stochastic perspective for latent sampling, where posterior inference is formulated as a diffusion process governed by a stochastic differential equation. This dynamic approach is particularly effective in capturing uncertainty evolution and temporally coherent structure across prostate MRI slices. Specifically, the latent variable $Z_t$ evolves via an Ornstein–Uhlenbeck-type process:
\begin{equation}
dZ_t = (\mu - Z_t) dt + \sigma\, dW_t,
\end{equation}
where $\mu$ and $\sigma$ are learned functions of the input image $x$, and $dW_t$ denotes the increment of a standard Wiener process. To align this evolution with the learned posterior from NF-Posterior, SDE-Girsanov employs Girsanov’s theorem to perform a measure transformation:
\begin{equation}
d\tilde{W}_t = dW_t + u_t dt, \quad u_t = \frac{\mu - Z_t}{\sigma},
\end{equation}
resulting in an equivalent SDE under a new measure $\mathbb{Q}$:
\begin{equation}
dZ_t = \sigma\, d\tilde{W}_t.
\end{equation}
The corresponding likelihood ratio between $\mathbb{Q}$ and the original measure $\mathbb{P}$ is expressed by the Radon–Nikodym derivative:
\begin{equation}
\frac{d\mathbb{Q}}{d\mathbb{P}} = \exp\left( -\frac{1}{2} \int_0^T \lambda_t^2 \, dt + \int_0^T \lambda_t\, dW_t \right),
\end{equation}
\begin{equation}
\quad \lambda_t = \frac{\mu - Z_t}{\sigma}.
\end{equation}
In implementation, the SDE is discretized with Euler–Maruyama steps:
\begin{equation}
Z_{t+\Delta t} = Z_t + (\mu - Z_t) \Delta t + \sigma \cdot \tilde{\varepsilon}_t,
\end{equation}
where $\tilde{\varepsilon}_t$ is a Radon–Nikodym-weighted Gaussian perturbation. This temporal evolution allows the latent code to transition gradually from the prior toward the posterior, embedding both anatomical priors and data-driven uncertainty. In prostate segmentation, SDE-Girsanov provides temporal smoothness and structure-consistent stochasticity across adjacent slices, improving robustness in regions with ambiguous tissue interfaces.

\subsection{Spatial Priors, Discrete Segmentation, and Bayesian Regularization}
\label{subsec:spatial-priors-bayes}

The model integrates spatial structural priors through Laplacian regularization applied to latent segmentation variables, which encourages smoothness and helps maintain anatomical boundary integrity. This is implemented by convolving latent shape and segmentation maps with a discrete Laplacian operator $\mathbf{D}_x$ that penalizes abrupt spatial variations, promoting spatial continuity.

To support end-to-end training with discrete segmentation maps, the Gumbel-Softmax reparameterization is employed. This technique provides a differentiable approximation to categorical sampling, enabling gradient-based optimization while preserving discrete label assignments during inference.

A pseudo-prior distribution $\pi$ is modeled on the categorical latent variables to introduce category-level regularization. This spatially varying distribution captures class prevalence and spatial arrangement patterns, guiding the posterior inference toward anatomically plausible segmentations and reducing spurious predictions.

The overall training loss combines supervised segmentation objectives, such as Dice and cross-entropy losses, with Bayesian regularization terms derived from the evidence lower bound (ELBO). The regularization terms include KL divergences between the variational posteriors and structured priors over latent variables representing appearance, shape, and segmentation logits. These KL terms are normalized by the number of spatial elements $N$ and weighted by a hyperparameter $\lambda_{\text{Bayes}}$ to balance their impact.

Formally, the loss is defined as:
\begin{equation}
\left\{
\begin{aligned}
\mathcal{L} &= \mathcal{L}_{\text{Dice+CE}} + \lambda_{\text{Bayes}} \cdot \frac{1}{N} \, \mathcal{L}_{\text{KL}}, \\
\mathcal{L}_{\text{KL}} &= 
\mathrm{KL}[q(\mathbf{m}) \| p(\mathbf{m})] + \mathcal{L}_{\text{KL}_x} + \mathcal{L}_{\text{KL}_z}, \\
\mathcal{L}_{\text{KL}_x} &= \mathrm{KL}[q(\mathbf{x}) \| p(\mathbf{x})], \\
\mathcal{L}_{\text{KL}_z} &= \mathrm{KL}[q(\mathbf{z}) \| p(\mathbf{z})].
\end{aligned}
\right.
\end{equation}

\section{Experiments}
\subsection{Datasets}
\label{datasets}
We utilized two public datasets for prostate segmentation: NCI-ISBI 2013 \cite{bloch2015nci} and PROMISE12 \cite{litjens2014evaluation}. The NCI-ISBI 2013 dataset includes 60 T2-weighted prostate MRI training volumes from multiple institutions with expert-annotated masks, capturing diverse anatomical and imaging variations. PROMISE12 contains 50 T2-weighted prostate MRI scans from various clinical centers acquired using different scanners and protocols, providing training labels while test labels are withheld for online evaluation; it is widely used to assess algorithm generalizability. All images were resampled to 0.36458×0.36458 mm spacing and histogram-clipped. The 3D volumes were converted into 2D slices, excluding those without the prostate, center-cropped to 384×384, and resized to 192×192. Preprocessing included Z-score normalization and random augmentations such as affine and elastic transformations and Gaussian noise to improve model robustness and generalization.

\subsection{Implementation Details}
\label{details}

All experiments were implemented in PyTorch using our proposed \textit{DeepBayesFlow} model and conducted on a single NVIDIA vGPU with 48GB of memory. Training was performed with a batch size of 8 for 1200 epochs. The Adam optimizer was used with an initial learning rate of 3e-4, which decayed once at epoch 1000, and a weight decay of 1e-4. To ensure reproducibility, the random seed was fixed at 42 and 4 worker threads were used for data loading. The model accepts single-channel input images and is optimized using a combined loss function, where the cross-entropy loss coefficient was set to 1.0 and the Bayesian loss coefficient to 100.0. During training, model checkpoints were regularly saved, and resuming from specific epochs was supported. For the Bayesian components, prior distributions were configured as follows: appearance mean with $\mu_0 = 0$ and $\sigma_0 = 1$; appearance standard deviation prior $\phi_\rho = 1 \times 10^{-6}$ and $\gamma_\rho = 2$; image boundary prior $\phi_\upsilon = 1 \times 10^{-8}$ and $\gamma_\upsilon = 2$; segmentation boundary prior $\phi_\omega = 1 \times 10^{-4}$ and $\gamma_\omega = 2$; and segmentation categorical prior with $\alpha_\pi = 2$ and $\beta_\pi = 2$.

\begin{table*}[ht]
\centering
\caption{Domain generalization performance across four datasets with RUNMC as source. The average is computed over all target domains.}
\label{tab:domain_generalization}
\begin{tabular}{l|cccc|c}
\toprule
\textbf{Method} & \textbf{RUNMC (Source)} & \textbf{BMC} & \textbf{BIDMC} & \textbf{HK} & \textbf{Avg. on Targets} \\
\midrule
ERM     & 83.8 & 73.7 & 17.7 & 68.5 & 53.3 \\
Cutout \cite{devries2017improved}  & 82.7 & 72.6 & 15.6 & 69.1 & 52.4 \\
IBN-Net \cite{pan2018two} & 84.0 & 78.4 & 42.8 & 76.0 & 65.7 \\
RandConv \cite{xu2020robust} & 84.0 & 75.3 & 16.6 & 30.8 & 40.9 \\
DSU \cite{li2022uncertainty}    & 84.1 & 78.7 & 31.0 & 73.9 & 61.2 \\
BayeSeg \cite{gao2023bayeseg} & \textbf{84.5} & 77.7 & 63.2 & 82.6 & 74.5 \\
\textbf{Ours}   & 82.4 & \textbf{81.1} & \textbf{74.7} & \textbf{83.7} & \textbf{79.8} \\
\bottomrule
\end{tabular}
\end{table*}

\begin{table*}[ht]
\centering
\caption{Domain generalization performance across four datasets with BMC as source. The average is computed over all target domains.}
\label{tab:domain_generalization_bmc}
\begin{tabular}{l|cccc|c}
\toprule
\textbf{Method} & \textbf{BMC (Source)} & \textbf{RUNMC} & \textbf{BIDMC} & \textbf{HK} & \textbf{Avg. on Targets} \\
\midrule
ERM     & 82.7 & 69.2 & 20.3 & 57.3 & 48.9 \\
Cutout \cite{devries2017improved}  & 81.6 & 68.1 & 18.1 & 58.6 & 48.2 \\
IBN-Net \cite{pan2018two} & 82.9 & 73.9 & 25.7 & 65.5 & 55.0 \\
RandConv \cite{xu2020robust} & 82.9 & 70.8 & 19.5 & 47.1 & 45.8 \\
DSU \cite{li2022uncertainty}    & 83.0 & \textbf{74.2} & 23.8 & 63.3 & 53.7 \\
BayeSeg \cite{gao2023bayeseg} & \textbf{83.4} & 73.2 & 27.6 & 69.2 & 56.6 \\
\textbf{Ours}   & 83.0 & 72.9 & \textbf{43.0} & \textbf{75.2} & \textbf{63.7} \\
\bottomrule
\end{tabular}
\end{table*}

\subsection{Results}
\label{results}

Table 1 and Table 2 report the domain generalization performance for prostate MRI segmentation, where models are trained on a single source domain (either RUNMC or BMC) and evaluated on three unseen target domains. This setup reflects practical clinical deployment scenarios in which models encounter significant distributional shifts due to differences in scanner hardware, imaging protocols, and patient anatomy.

In the case where RUNMC is used as the source domain, the proposed method achieves the best performance across all target domains. While baseline methods such as ERM and Cutout exhibit substantial drops in performance, particularly on BIDMC, the proposed method obtains a Dice score of 74.7 on BIDMC and 83.7 on HK, resulting in the highest average performance of 79.8 across all target domains. Compared to BayeSeg, the proposed method yields consistent improvements, especially on more challenging datasets such as BIDMC, where the performance increases by 11.5 points.

When BMC is used as the source domain, the proposed method maintains its superiority. It achieves a Dice score of 43.0 on BIDMC and 75.2 on HK, with an overall average of 63.7 across the three target domains. This is significantly higher than the average score of 56.6 obtained by BayeSeg, again demonstrating enhanced robustness to domain shifts and improved adaptability in out-of-distribution settings.

Robust generalization is essential for prostate MRI segmentation in real-world clinical applications. The proposed method shows stable and accurate segmentation across heterogeneous domains without requiring retraining or fine-tuning. This level of robustness is crucial for safe and effective clinical deployment across hospitals with varying imaging protocols.

The visualization results of the proposed DeepBayesFlow model against other state-of-the-arts are shown in Figure 2. Figure 2 presents the visual segmentation results of various prostate MRI segmentation models across three public datasets: BIDMC (left), BMC (middle), and HK (right). Each row corresponds to a different segmentation method, including ERM, Cutout, IBN-Net, RandConv, DSU, BayeSeg, and DeepBayesFlow. Each column displays representative slices from the respective datasets. The segmentation outcomes are color-coded as follows: red regions indicate false positives (predicted as foreground but actually background), green regions indicate false negatives (missed foreground regions), and purple regions indicate true positives (correctly predicted foreground).

From Figure 2, it is evident that conventional methods such as ERM and Cutout exhibit large false positive and false negative regions across datasets, revealing poor generalization ability. IBN-Net and RandConv show improvements but still suffer from discontinuous or incomplete structures. BayeSeg demonstrates more robust performance, reducing both false positives and false negatives to some extent. However, in certain cases, red and green regions remain, suggesting limitations in boundary precision and sensitivity.

In contrast, DeepBayesFlow produces cleaner segmentation results with minimal red or green regions and more consistent purple regions, indicating high prediction accuracy and reliable boundary delineation. Compared to BayeSeg, DeepBayesFlow shows superior performance in terms of structural continuity, shape consistency, and domain generalization. This improvement is particularly noticeable in the HK dataset, where domain shifts are more pronounced due to differences in imaging modalities and anatomical variations. DeepBayesFlow maintains accurate and stable segmentation across these challenging conditions.

\begin{table*}[ht]
\centering
\caption{Ablation study of different version combinations with RUNMC as source domain. Each version includes/excludes components NF-Posterior, NCVI, and SDE-Girsanov. The average is computed over all target domains.}
\label{tab:ablation_versions}
\begin{tabular}{l|ccc|c|cccc|c}
\toprule
\textbf{Version} & \textbf{NF-Posterior} & \textbf{NCVI} & \textbf{SDE-Girsanov} & \textbf{RUNMC (Source)} & \textbf{BMC} & \textbf{BIDMC} & \textbf{HK} & \textbf{Avg. on Targets} \\
\midrule
Ver 1 & $\times$ & $\times$ & $\times$ & 84.5 & 77.7 & 63.2 & 82.6 & 74.5 \\
Ver 2 & $\times$ & $\checkmark$ & $\checkmark$ & 80.6 & 80.4 & 70.4 & 82.7 & 77.8 \\
Ver 3 & $\checkmark$ & $\times$ & $\checkmark$ & \textbf{86.1} & 77.5 & 73.8 & \textbf{83.7} & 78.3 \\
Ver 4 & $\checkmark$ & $\checkmark$ & $\times$ & 81.5 & 79.9 & 70.6 & 81.4 & 77.3 \\
\textbf{Ver 5} & $\checkmark$ & $\checkmark$ & $\checkmark$ & 82.4 & \textbf{81.1} & \textbf{74.7} & \textbf{83.7} & \textbf{79.8} \\
\bottomrule
\end{tabular}
\end{table*}

\begin{table*}[ht]
\centering
\caption{Ablation study of different version combinations with BMC as source domain. Each version includes/excludes components NF-Posterior, NCVI, and SDE-Girsanov. The average is computed over all target domains.}
\label{tab:ablation_versions_bmc}
\begin{tabular}{l|ccc|c|cccc|c}
\toprule
\textbf{Version} & \textbf{NF-Posterior} & \textbf{NCVI} & \textbf{SDE-Girsanov} & \textbf{BMC (Source)} & \textbf{RUNMC} & \textbf{BIDMC} & \textbf{HK} & \textbf{Avg. on Targets} \\
\midrule
Ver 1 & $\times$ & $\times$ & $\times$ & 83.4 & 73.2 & 27.6 & 69.2 & 56.6 \\
Ver 2 & $\times$ & $\checkmark$ & $\checkmark$ & 83.4 & 66.0 & 40.2 & 63.9 & 56.7 \\
Ver 3 & $\checkmark$ & $\times$ & $\checkmark$ & \textbf{85.5} & \textbf{75.6} & 36.5 & 63.9 & 58.6 \\
Ver 4 & $\checkmark$ & $\checkmark$ & $\times$ & 83.6 & 72.0 & 36.8 & 62.9 & 57.2 \\
\textbf{Ver 5} & $\checkmark$ & $\checkmark$ & $\checkmark$ & 83.0 & 72.9 & \textbf{43.0} & \textbf{75.2} & \textbf{63.7} \\
\bottomrule
\end{tabular}
\end{table*}

\begin{figure*}[htbp]
\centering
\includegraphics[width=\textwidth]{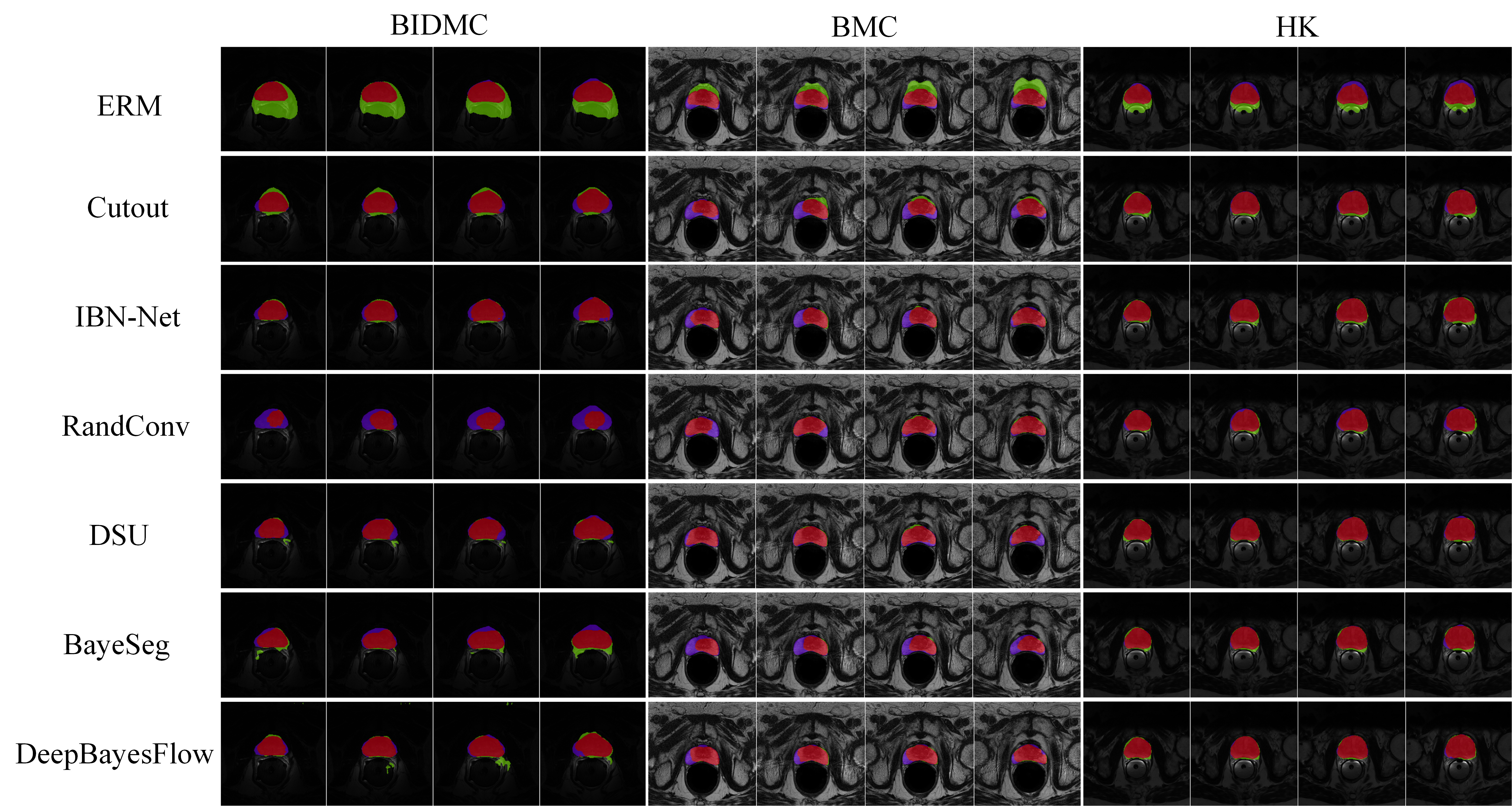}
\caption{The visualization results of the proposed DeepBayesFlow model against other state-of-the-arts.}
\label{fig:tu1}
\end{figure*}

\subsection{Discussion}
\label{discussion}

Table 3 and Table 4 report the results of an ablation study designed to assess the individual and combined contributions of three key components in the model: NF-Posterior, NCVI, and SDE-Girsanov. Experiments are conducted under two different source domain settings, RUNMC and BMC, to evaluate the model's generalization ability across unseen target domains.

When RUNMC is used as the source domain, the base version without any of the three components (Ver 1) achieves an average Dice score of 74.5 across the target domains. Adding NCVI and SDE-Girsanov (Ver 2) leads to a notable improvement, especially on BIDMC, raising the average score to 77.8. This suggests that the combination of noise-contrastive variational inference and stochastic perturbation improves robustness against domain shift and helps mitigate false negatives in challenging regions. Ver 3, which includes NF-Posterior and SDE-Girsanov, achieves further gains, highlighting the effectiveness of using a flexible posterior to better model anatomical uncertainty. Ver 4, combining NF-Posterior and NCVI, also improves the baseline but is slightly less effective than the combination in Ver 3. The full version (Ver 5), which incorporates all three components, consistently outperforms all other configurations, achieving an average target-domain Dice score of 79.8. This demonstrates the complementary nature of the components and their joint benefit in improving segmentation performance.

Similar trends are observed when BMC is used as the source domain. The base model (Ver 1) reaches an average Dice of 56.6. Adding NCVI and SDE-Girsanov (Ver 2) slightly improves performance to 56.7, and the inclusion of NF-Posterior in Ver 3 and Ver 4 provides further gains. The complete model (Ver 5) again achieves the highest scores on all target domains, with an average Dice of 63.7, indicating that all components contribute to generalization, especially in the most challenging cases such as BIDMC.

Each module addresses a specific challenge inherent in prostate MRI segmentation. NF-Posterior enhances the model's ability to represent complex and diverse anatomical distributions, which is crucial for capturing inter-patient variability in prostate shape and appearance. NCVI encourages better uncertainty modeling by introducing contrastive learning in the latent space, which helps the model distinguish between reliable and ambiguous predictions under distribution shift. SDE-Girsanov introduces controlled stochasticity in the inference process, which stabilizes predictions in the presence of noise and helps the model adapt to low-contrast boundaries, common in real clinical scans.

These three components work in a complementary manner: while NF-Posterior improves representation capacity, NCVI regularizes the latent space, and SDE-Girsanov introduces robustness to temporal and spatial perturbations. The result is a segmentation model that better generalizes across domains, reduces false positives and false negatives, and maintains anatomical consistency. In clinical practice, this translates to more reliable delineation of the prostate across patients and institutions, which is essential for accurate cancer detection, treatment planning, and follow-up evaluation.

\section{Conclusion}

In this work, we presented \textit{DeepBayesFlow}, a novel Bayesian segmentation framework designed to address the key challenges of prostate MRI segmentation, including anatomical variability, boundary ambiguity, and domain shifts across imaging protocols. By integrating a learnable normalizing flow-based posterior (NF-Posterior), a non-conjugate variational inference strategy (NCVI), and a time-continuous SDE-based latent diffusion mechanism with Girsanov transformation (SDE-Girsanov), our model achieves expressive posterior modeling, robust uncertainty estimation, and strong generalization capability. These components work synergistically to capture domain-invariant structural features while adapting flexibly to distribution-specific variations. Comprehensive experiments on multi-domain prostate segmentation datasets demonstrate that DeepBayesFlow outperforms state-of-the-art methods in both accuracy and robustness. In future work, we plan to extend this framework to broader segmentation tasks and investigate the integration of domain-adaptive priors for further enhancing interpretability and cross-domain reliability.

\bibliography{aaai2026}

@article{liu2021review,
  author    = {Liu, Xiangbin and Zhang, Yong and Zhang, Ying and Wang, Haitao and Wang, Yimin},
  title     = {A Review of Deep-Learning-Based Medical Image Segmentation Methods},
  journal   = {Sustainability},
  volume    = {13},
  number    = {3},
  pages     = {1224},
  year      = {2021},
  doi       = {10.3390/su13031224},
  issn      = {2071-1050},
  publisher = {MDPI},
  url       = {https://www.mdpi.com/2071-1050/13/3/1224}
}

@article{conze2023trends,
  author    = {Conze, Pierre-Henri and Jalal, Omar and Robert, Camille and Smistad, Erik and Kerautret, Bertrand and Cervenansky, Frédéric and Quellec, Gwénolé},
  title     = {Current and Emerging Trends in Medical Image Segmentation with Deep Learning},
  journal   = {IEEE Transactions on Radiation and Plasma Medical Sciences},
  volume    = {7},
  number    = {6},
  pages     = {545--569},
  year      = {2023},
  doi       = {10.1109/TRPMS.2023.3266647},
  issn      = {2469-7311},
  publisher = {IEEE},
  url       = {https://ieeexplore.ieee.org/document/10114486}
}

@article{wang2023twostage,
  author    = {Wang, Zixuan and Ma, Hong and Liu, Zhaoyang and Zhang, Kun and Li, Lin},
  title     = {A Two-Stage CNN Method for MRI Image Segmentation of Prostate with Lesion},
  journal   = {Biomedical Signal Processing and Control},
  volume    = {82},
  pages     = {104610},
  year      = {2023},
  doi       = {10.1016/j.bspc.2022.104610},
  issn      = {1746-8094},
  publisher = {Elsevier},
  url       = {https://www.sciencedirect.com/science/article/pii/S1746809422009254}
}

@article{chen2021alexnet,
  author    = {Chen, Jun and Zhang, Chunlan and Qian, Yichun and Xie, Yujun and Wang, Jincao},
  title     = {Medical Image Segmentation and Reconstruction of Prostate Tumor Based on 3D AlexNet},
  journal   = {Computer Methods and Programs in Biomedicine},
  volume    = {200},
  pages     = {105878},
  year      = {2021},
  doi       = {10.1016/j.cmpb.2020.105878},
  issn      = {0169-2607},
  publisher = {Elsevier},
  url       = {https://www.sciencedirect.com/science/article/pii/S0169260720315804}
}

@article{jia2022synergic,
  author    = {Jia, Haozhe and Gao, Yujia and Wang, Ruixuan and Wang, Zhiguo and Guo, Yujie and Gao, Yang and Liu, Jiang},
  title     = {Learning Multi-Scale Synergic Discriminative Features for Prostate Image Segmentation},
  journal   = {Pattern Recognition},
  volume    = {126},
  pages     = {108556},
  year      = {2022},
  doi       = {10.1016/j.patcog.2022.108556},
  issn      = {0031-3203},
  publisher = {Elsevier},
  url       = {https://www.sciencedirect.com/science/article/pii/S0031320322000082}
}

@article{bhandary2023unet,
  author    = {Bhandary, Shrajan and Krishnamurthi, Ganapathy and Deshpande, Sachin and Palaniappan, Kannappan},
  title     = {Investigation and Benchmarking of U-Nets on Prostate Segmentation Tasks},
  journal   = {Computerized Medical Imaging and Graphics},
  volume    = {107},
  pages     = {102241},
  year      = {2023},
  doi       = {10.1016/j.compmedimag.2023.102241},
  issn      = {0895-6111},
  publisher = {Elsevier},
  url       = {https://www.sciencedirect.com/science/article/pii/S0895611123000497}
}

@article{jiang2023ultrasound,
  author    = {Jiang, Jingang and Chen, Wenyao and Bai, Xiangyu and Qin, Jing and Cheng, Heng-Da and Jiang, Tianzi},
  title     = {Segmentation of Prostate Ultrasound Images: The State of the Art and the Future Directions of Segmentation Algorithms},
  journal   = {Artificial Intelligence Review},
  volume    = {56},
  number    = {1},
  pages     = {615--651},
  year      = {2023},
  doi       = {10.1007/s10462-022-10214-4},
  issn      = {0269-2821},
  publisher = {Springer},
  url       = {https://link.springer.com/article/10.1007/s10462-022-10214-4}
}

@article{sammouda2021kmeans,
  author    = {Sammouda, Rachid and El-Zaart, Ali},
  title     = {An Optimized Approach for Prostate Image Segmentation Using K-Means Clustering Algorithm with Elbow Method},
  journal   = {Computational Intelligence and Neuroscience},
  volume    = {2021},
  number    = {1},
  pages     = {4553832},
  year      = {2021},
  doi       = {10.1155/2021/4553832},
  issn      = {1687-5265},
  publisher = {Hindawi},
  url       = {https://www.hindawi.com/journals/cin/2021/4553832/}
}

@article{orlando2022ultrasound,
  author    = {Orlando, Nathan and Alqahtani, Sultan and Almekkawi, Ahmed and Abolmaesumi, Purang and Salcudean, Septimiu E. and Podder, Tania K. and Fenster, Aaron},
  title     = {Effect of Dataset Size, Image Quality, and Image Type on Deep Learning-Based Automatic Prostate Segmentation in 3D Ultrasound},
  journal   = {Physics in Medicine \& Biology},
  volume    = {67},
  number    = {7},
  pages     = {074002},
  year      = {2022},
  doi       = {10.1088/1361-6560/ac50d4},
  issn      = {1361-6560},
  publisher = {IOP Publishing},
  url       = {https://iopscience.iop.org/article/10.1088/1361-6560/ac50d4}
}

@article{wang2023dsunet,
  author    = {Wang, Xinyu and Wang, Yaping and Li, Jian and Zhang, Yujie and Song, Tao},
  title     = {Prostate Ultrasound Image Segmentation Based on DSU-Net},
  journal   = {Biomedicines},
  volume    = {11},
  number    = {3},
  pages     = {646},
  year      = {2023},
  doi       = {10.3390/biomedicines11030646},
  issn      = {2227-9059},
  publisher = {MDPI},
  url       = {https://www.mdpi.com/2227-9059/11/3/646}
}

@article{zaridis2023region,
  author    = {Zaridis, Dimitrios I. and Maros, Tamas N. and Stoll, Sebastian and Khalvati, Farzad and M{\"u}ller, Henning and Vangelov, Bozhidar and Eklund, Anders and S{\"o}rensen, Thomas S.},
  title     = {Region-Adaptive Magnetic Resonance Image Enhancement for Improving CNN-Based Segmentation of the Prostate and Prostatic Zones},
  journal   = {Scientific Reports},
  volume    = {13},
  number    = {1},
  pages     = {714},
  year      = {2023},
  doi       = {10.1038/s41598-023-27884-1},
  issn      = {2045-2322},
  publisher = {Nature Publishing Group},
  url       = {https://www.nature.com/articles/s41598-023-27884-1}
}

@article{yoon2024domain,
  author    = {Yoon, Jee Seok and Lee, Min Ju and Kim, Jin Woo and Park, Soo Min and Choi, Yong Jun and Kim, Hyun Soo and Lee, Kyoung Mu},
  title     = {Domain Generalization for Medical Image Analysis: A Review},
  journal   = {Proceedings of the IEEE},
  year      = {2024},
  doi       = {10.1109/JPROC.2024.XXXXXXX},
  publisher = {IEEE},
  note      = {To appear}
}

@article{ouyang2022causality,
  author    = {Ouyang, Cheng and Yuan, Jiang and Zhu, Yubo and Li, Yingda and Wang, Jiyang and Zhang, Jian and Wu, Jun},
  title     = {Causality-Inspired Single-Source Domain Generalization for Medical Image Segmentation},
  journal   = {IEEE Transactions on Medical Imaging},
  volume    = {42},
  number    = {4},
  pages     = {1095--1106},
  year      = {2022},
  doi       = {10.1109/TMI.2022.3148881},
  issn      = {0278-0062},
  publisher = {IEEE},
  url       = {https://ieeexplore.ieee.org/document/9734656}
}

@article{hu2022domain,
  author    = {Hu, Shishuai and Yuan, Jiang and Wu, Yuhui and Zhao, Ming and Zhang, Haifeng and Tian, Qi},
  title     = {Domain and Content Adaptive Convolution Based Multi-Source Domain Generalization for Medical Image Segmentation},
  journal   = {IEEE Transactions on Medical Imaging},
  volume    = {42},
  number    = {1},
  pages     = {233--244},
  year      = {2022},
  doi       = {10.1109/TMI.2022.3145973},
  issn      = {0278-0062},
  publisher = {IEEE},
  url       = {https://ieeexplore.ieee.org/document/9731938}
}

@inproceedings{su2023rethinking,
  author    = {Su, Zixian and Zhu, Ming and Wang, Zhaoyang and Liu, Bin and Xu, Hongzhi and Yu, Meng and Lu, Hui},
  title     = {Rethinking Data Augmentation for Single-Source Domain Generalization in Medical Image Segmentation},
  booktitle = {Proceedings of the AAAI Conference on Artificial Intelligence},
  volume    = {37},
  number    = {2},
  year      = {2023},
  pages     = {},
  publisher = {AAAI Press},
  url       = {https://ojs.aaai.org/index.php/AAAI/article/view/26088}
}

@inproceedings{liu2021feddg,
  author    = {Liu, Quande and Chen, Yuhang and Zhou, Jianfei and Zheng, Wenming and Cai, Weidong},
  title     = {FedDG: Federated Domain Generalization on Medical Image Segmentation via Episodic Learning in Continuous Frequency Space},
  booktitle = {Proceedings of the IEEE/CVF Conference on Computer Vision and Pattern Recognition (CVPR)},
  year      = {2021},
  pages     = {},
  publisher = {IEEE},
  url       = {https://openaccess.thecvf.com/content/CVPR2021/html/Liu_FedDG_Federated_Domain_Generalization_on_Medical_Image_Segmentation_via_Episodic_Learning_in_CVPR_2021_paper.html}
}

@inproceedings{xu2022adversarial,
  author    = {Xu, Yanwu and Huang, Kai and Chen, Haoran and Liu, Wei and Yan, Pengcheng and Li, Ke and Yu, Dong and Shao, Ling},
  title     = {Adversarial Consistency for Single Domain Generalization in Medical Image Segmentation},
  booktitle = {Medical Image Computing and Computer-Assisted Intervention -- MICCAI 2022},
  year      = {2022},
  publisher = {Springer Nature Switzerland},
  address   = {Cham},
  pages     = {300--310},
  doi       = {10.1007/978-3-031-16443-9\_29},
  url       = {https://link.springer.com/chapter/10.1007/978-3-031-16443-9_29}
}

@inproceedings{yao2022enhancing,
  author    = {Yao, Huifeng and Hu, Xiaowei and Li, Xiaomeng},
  title     = {Enhancing Pseudo Label Quality for Semi-Supervised Domain-Generalized Medical Image Segmentation},
  booktitle = {Proceedings of the AAAI Conference on Artificial Intelligence},
  volume    = {36},
  number    = {3},
  year      = {2022},
  publisher = {AAAI Press},
  url       = {https://ojs.aaai.org/index.php/AAAI/article/view/18853}
}

@inproceedings{gao2024desam,
  author    = {Gao, Yifan and Zhang, Peng and Liu, Heng and Wang, Lei and Chen, Yu and Li, Jian and Zhou, Wei},
  title     = {Desam: Decoupled Segment Anything Model for Generalizable Medical Image Segmentation},
  booktitle = {Medical Image Computing and Computer-Assisted Intervention -- MICCAI 2024},
  year      = {2024},
  publisher = {Springer Nature Switzerland},
  address   = {Cham},
  pages     = {},
  doi       = {},
  url       = {}
}

@article{lyu2022aadg,
  author    = {Lyu, Junyan and Guo, Yu and Wang, Xiaosong and Peng, Chao and Zhao, Shuyang and Li, Qiang and Wang, Lei},
  title     = {AADG: Automatic Augmentation for Domain Generalization on Retinal Image Segmentation},
  journal   = {IEEE Transactions on Medical Imaging},
  volume    = {41},
  number    = {12},
  pages     = {3699--3711},
  year      = {2022},
  doi       = {10.1109/TMI.2022.3190309},
  issn      = {0278-0062},
  publisher = {IEEE},
  url       = {https://ieeexplore.ieee.org/document/9742534}
}

@article{li2022domain,
  author    = {Li, Chenxin and Zhang, Rui and Wang, Sheng and Xu, Wen and Lu, Meng and Chen, Yujie and Zhang, Mingliang},
  title     = {Domain Generalization on Medical Imaging Classification Using Episodic Training with Task Augmentation},
  journal   = {Computers in Biology and Medicine},
  volume    = {141},
  pages     = {105144},
  year      = {2022},
  doi       = {10.1016/j.compbiomed.2021.105144},
  issn      = {0010-4825},
  publisher = {Elsevier},
  url       = {https://www.sciencedirect.com/science/article/pii/S0010482521006196}
}

@inproceedings{wang2022rethinking,
  author    = {Wang, Jianfeng and Lukasiewicz, Thomas},
  title     = {Rethinking Bayesian Deep Learning Methods for Semi-Supervised Volumetric Medical Image Segmentation},
  booktitle = {Proceedings of the IEEE/CVF Conference on Computer Vision and Pattern Recognition (CVPR)},
  year      = {2022},
  publisher = {IEEE},
  url       = {https://openaccess.thecvf.com/content/CVPR2022/html/Wang_Rethinking_Bayesian_Deep_Learning_Methods_for_Semi-Supervised_Volumetric_Medical_Image_Segmentation_CVPR_2022_paper.html}
}

@article{gao2023bayeseg,
  author    = {Gao, Shangqi and Hu, Shishuai and Li, Wentao and Xu, Xiao and Zhu, Sheng and Jiang, Hao and Shen, Dinggang and Zhang, Daguang},
  title     = {BayeSeg: Bayesian Modeling for Medical Image Segmentation with Interpretable Generalizability},
  journal   = {Medical Image Analysis},
  volume    = {89},
  pages     = {102889},
  year      = {2023},
  doi       = {10.1016/j.media.2023.102889},
  issn      = {1361-8415},
  publisher = {Elsevier},
  url       = {https://www.sciencedirect.com/science/article/pii/S1361841523000406}
}

@inproceedings{zhao2022efficient,
  author    = {Zhao, Yidong and Huang, Wen and Wang, Lei and Ma, Kun and Zhan, Xuan and Zhang, Lei and Xu, Xiaofeng and Shen, Dinggang},
  title     = {Efficient Bayesian Uncertainty Estimation for NNU-Net},
  booktitle = {Medical Image Computing and Computer-Assisted Intervention -- MICCAI 2022},
  year      = {2022},
  publisher = {Springer Nature Switzerland},
  address   = {Cham},
  pages     = {122--131},
  doi       = {10.1007/978-3-031-16443-9\_12},
  url       = {https://link.springer.com/chapter/10.1007/978-3-031-16443-9_12}
}

@article{bloch2015nci,
  author    = {Bloch, Nicholas and Madabhushi, Anant and Feldman, Mark and Choyke, Peter and Kikinis, Ron and Tannenbaum, Allen and Giger, Maryellen and Summers, Ronald},
  title     = {NCI-ISBI 2013 Challenge: Automated Segmentation of Prostate Structures},
  journal   = {The Cancer Imaging Archive},
  volume    = {370},
  number    = {6},
  pages     = {5},
  year      = {2015},
  publisher = {},
  url       = {}
}

@article{litjens2014evaluation,
  author    = {Litjens, Geert and Toth, Reka and van de Ven, Wim and Hoeks, Cas and Kerkstra, Sjoerd and van Ginneken, Bram and Karssemeijer, Nico and Huisman, Henk},
  title     = {Evaluation of Prostate Segmentation Algorithms for MRI: The PROMISE12 Challenge},
  journal   = {Medical Image Analysis},
  volume    = {18},
  number    = {2},
  pages     = {359--373},
  year      = {2014},
  doi       = {10.1016/j.media.2013.12.002},
  issn      = {1361-8415},
  publisher = {Elsevier},
  url       = {https://www.sciencedirect.com/science/article/pii/S1361841513001605}
}

@misc{devries2017improved,
  author       = {DeVries, Terrance and Taylor, Graham W.},
  title        = {Improved Regularization of Convolutional Neural Networks with Cutout},
  year         = {2017},
  eprint       = {1708.04552},
  archivePrefix= {arXiv},
  primaryClass = {cs.CV},
  url          = {https://arxiv.org/abs/1708.04552}
}

@inproceedings{pan2018two,
  author    = {Pan, Xingang and Jiang, Bo and Yang, Yi and Shi, Jian and Wang, Xiaojuan and Lu, Jiaya},
  title     = {Two at Once: Enhancing Learning and Generalization Capacities via IBN-Net},
  booktitle = {Proceedings of the European Conference on Computer Vision (ECCV)},
  year      = {2018},
  pages     = {464--479},
  publisher = {Springer},
  address   = {},
  doi       = {10.1007/978-3-030-01246-5_28},
  url       = {https://doi.org/10.1007/978-3-030-01246-5_28}
}

@misc{xu2020robust,
  author       = {Xu, Zhenlin and Liu, Zhiyu and Huang, Yongwei and Dai, Bo and Li, Chen and Song, Yongxin and Li, Weiming},
  title        = {Robust and Generalizable Visual Representation Learning via Random Convolutions},
  year         = {2020},
  eprint       = {2007.13003},
  archivePrefix= {arXiv},
  primaryClass = {cs.CV},
  url          = {https://arxiv.org/abs/2007.13003}
}

@misc{li2022uncertainty,
  author       = {Li, Xiaotong and Huang, Yixuan and Xu, Yuchen and Liu, Zhi and Gu, Songchun and Darrell, Trevor and Rohrbach, Marcus},
  title        = {Uncertainty Modeling for Out-of-Distribution Generalization},
  year         = {2022},
  eprint       = {2202.03958},
  archivePrefix= {arXiv},
  primaryClass = {cs.LG},
  url          = {https://arxiv.org/abs/2202.03958}
}

@article{azad2024medical,
  author    = {Azad, Reza and Devi, Dwarikanath Mahapatra and Merhof, Dorit and Bhatia, Kuldeep Kumar and Zhou, Shuo and Asadi-Aghbolaghi, Mahmood},
  title     = {Medical Image Segmentation Review: The Success of U-Net},
  journal   = {IEEE Transactions on Pattern Analysis and Machine Intelligence},
  year      = {2024},
  doi       = {10.1109/TPAMI.2024.3371822},
  publisher = {IEEE},
  url       = {https://ieeexplore.ieee.org/document/10432604}
}

@article{ramesh2021review,
  author    = {Ramesh, K. K. D. and Nandhini, K. and Gomathi, R. and Kumar, M. S.},
  title     = {A Review of Medical Image Segmentation Algorithms},
  journal   = {EAI Endorsed Transactions on Pervasive Health and Technology},
  volume    = {7},
  number    = {27},
  year      = {2021},
  doi       = {10.4108/eai.13-7-2018.164261},
  publisher = {European Alliance for Innovation (EAI)},
  url       = {https://publications.eai.eu/index.php/phat/article/view/1642}
}

@inproceedings{butoi2023universeg,
  author    = {Butoi, Victor Ion and Chang, Jin-Hwa and Lee, Jaesik and Bengio, Yoshua and Paudel, Danda and Rother, Carsten},
  title     = {Universeg: Universal Medical Image Segmentation},
  booktitle = {Proceedings of the IEEE/CVF International Conference on Computer Vision (ICCV)},
  year      = {2023},
  publisher = {IEEE},
  pages     = {21554--21564},
  url       = {https://openaccess.thecvf.com/content/ICCV2023/html/Butoi_Universeg_Universal_Medical_Image_Segmentation_ICCV_2023_paper.html}
}

@article{wu2025medical,
  author    = {Wu, Junde and Luo, Lequan and Huang, Huaiqiu and Wu, Jie and Wang, Xiang and Wang, Yizhou and Li, Hongsheng},
  title     = {Medical SAM Adapter: Adapting Segment Anything Model for Medical Image Segmentation},
  journal   = {Medical Image Analysis},
  volume    = {102},
  pages     = {103547},
  year      = {2025},
  doi       = {10.1016/j.media.2024.103547},
  publisher = {Elsevier},
  url       = {https://www.sciencedirect.com/science/article/pii/S1361841524001520}
}

@article{zhang2024segment,
  author    = {Zhang, Yichi and Shen, Zhenrong and Jiao, Rushi},
  title     = {Segment Anything Model for Medical Image Segmentation: Current Applications and Future Directions},
  journal   = {Computers in Biology and Medicine},
  volume    = {171},
  pages     = {108238},
  year      = {2024},
  doi       = {10.1016/j.compbiomed.2024.108238},
  publisher = {Elsevier},
  url       = {https://www.sciencedirect.com/science/article/pii/S0010482524002143}
}

@inproceedings{salpea2022medical,
  author    = {Salpea, Natalia and Tzouveli, Paraskevi and Kollias, Dimitrios},
  title     = {Medical Image Segmentation: A Review of Modern Architectures},
  booktitle = {European Conference on Computer Vision (ECCV)},
  year      = {2022},
  publisher = {Springer Nature Switzerland},
  address   = {Cham},
  pages     = {},
  doi       = {},
  url       = {}
}

@misc{he2023accuracy,
  author       = {He, Sheng and Liu, Hongwei and Zhu, Wenhao and Dong, Mengya and Zhang, Xiaotong and Li, Zhiyong and Xu, Maosong},
  title        = {Accuracy of Segment-Anything Model (SAM) in Medical Image Segmentation Tasks},
  year         = {2023},
  eprint       = {2304.05396},
  archivePrefix= {arXiv},
  primaryClass = {eess.IV},
  url          = {https://arxiv.org/abs/2304.05396}
}

@article{rayed2024deep,
  author    = {Rayed, Md Eshmam and Morshed, Md Nazmus Sakib and Kamal, Md Masudur Rahman and Momen, Saifur Rahman and Ahmed, Khandakar Nafiz Towfiq and Amin, Md Rezwanul},
  title     = {Deep Learning for Medical Image Segmentation: State-of-the-Art Advancements and Challenges},
  journal   = {Informatics in Medicine Unlocked},
  volume    = {47},
  pages     = {101504},
  year      = {2024},
  doi       = {10.1016/j.imu.2024.101504},
  publisher = {Elsevier},
  url       = {https://www.sciencedirect.com/science/article/pii/S2352914824000604}
}

@inproceedings{valanarasu2022unext,
  author    = {Valanarasu, Jeya Maria Jose and Patel, Vishal M.},
  title     = {UNeXt: MLP-Based Rapid Medical Image Segmentation Network},
  booktitle = {International Conference on Medical Image Computing and Computer-Assisted Intervention (MICCAI)},
  year      = {2022},
  publisher = {Springer Nature Switzerland},
  address   = {Cham},
  pages     = {96--106},
  doi       = {10.1007/978-3-031-16443-4_10},
  url       = {https://link.springer.com/chapter/10.1007/978-3-031-16443-4_10}
}

@inproceedings{xing2024segmamba,
  author    = {Xing, Zhaohu and Zhang, Yuan and Chen, Kun and Liu, Yuxin and Wang, Xiaoli and Zhang, Rui and Li, Ming and Wang, Xinyu},
  title     = {SegMamba: Long-Range Sequential Modeling Mamba for 3D Medical Image Segmentation},
  booktitle = {International Conference on Medical Image Computing and Computer-Assisted Intervention (MICCAI)},
  year      = {2024},
  publisher = {Springer Nature Switzerland},
  address   = {Cham},
  pages     = {},
  doi       = {},
  url       = {}
}

@article{chen2024ma,
  author    = {Chen, Cheng and Li, Xiaotong and Wang, Jia and Zhang, Hong and Wang, Yibin and Li, Shun and Zhou, Shuo and Wang, Dinggang},
  title     = {Ma-SAM: Modality-Agnostic SAM Adaptation for 3D Medical Image Segmentation},
  journal   = {Medical Image Analysis},
  volume    = {98},
  pages     = {103310},
  year      = {2024},
  doi       = {10.1016/j.media.2024.103310},
  publisher = {Elsevier},
  url       = {https://www.sciencedirect.com/science/article/pii/S1361841523003245}
}

@inproceedings{roy2023mednext,
  author    = {Roy, Saikat and Valanarasu, Jeya Maria Jose and Oza, Purva and Hacihaliloglu, Ilker and Patel, Vishal M.},
  title     = {MedNeXt: Transformer-Driven Scaling of ConvNets for Medical Image Segmentation},
  booktitle = {International Conference on Medical Image Computing and Computer-Assisted Intervention (MICCAI)},
  year      = {2023},
  publisher = {Springer Nature Switzerland},
  address   = {Cham},
  pages     = {},
  doi       = {},
  url       = {}
}

@article{xiao2023transformers,
  author    = {Xiao, Hanguang and Sun, Meng and Han, Qian and Fang, Zhixiang and Gao, Jing and Yang, Guangyu and Gao, Guangyao and Feng, Dagan and Li, Shuo},
  title     = {Transformers in Medical Image Segmentation: A Review},
  journal   = {Biomedical Signal Processing and Control},
  volume    = {84},
  pages     = {104791},
  year      = {2023},
  doi       = {10.1016/j.bspc.2023.104791},
  publisher = {Elsevier},
  url       = {https://www.sciencedirect.com/science/article/pii/S1746809423002327}
}

@article{muller2022towards,
  author    = {Müller, Dominik and Soto-Rey, Iñaki and Kramer, Frank},
  title     = {Towards a Guideline for Evaluation Metrics in Medical Image Segmentation},
  journal   = {BMC Research Notes},
  volume    = {15},
  number    = {1},
  pages     = {210},
  year      = {2022},
  doi       = {10.1186/s13104-022-06125-6},
  publisher = {BioMed Central},
  url       = {https://bmcresnotes.biomedcentral.com/articles/10.1186/s13104-022-06125-6}
}

@article{chen2024transunet,
  author    = {Chen, Jieneng and Lu, Yong and Yu, Qihang and Luo, Xiangde and Adeli, Ehsan and Wang, Yuyin and Lu, Lin and Yuille, Alan L. and Zhou, Yuyin},
  title     = {TransUNet: Rethinking the U-Net Architecture Design for Medical Image Segmentation through the Lens of Transformers},
  journal   = {Medical Image Analysis},
  volume    = {97},
  pages     = {103280},
  year      = {2024},
  doi       = {10.1016/j.media.2023.103280},
  publisher = {Elsevier},
  url       = {https://www.sciencedirect.com/science/article/pii/S1361841523003383}
}

@misc{zhang2023customized,
  author       = {Zhang, Kaidong and Liu, Dong},
  title        = {Customized Segment Anything Model for Medical Image Segmentation},
  year         = {2023},
  eprint       = {2304.13785},
  archivePrefix= {arXiv},
  primaryClass = {cs.CV},
  url          = {https://arxiv.org/abs/2304.13785}
}

@inproceedings{isensee2024nnu,
  author    = {Isensee, Fabian and Jaeger, Patrick F. and Kohl, Simon A. A. and Petersen, Jens and Maier-Hein, Klaus H.},
  title     = {nnU-Net Revisited: A Call for Rigorous Validation in 3D Medical Image Segmentation},
  booktitle = {International Conference on Medical Image Computing and Computer-Assisted Intervention (MICCAI)},
  year      = {2024},
  publisher = {Springer Nature Switzerland},
  address   = {Cham},
  pages     = {},
  doi       = {},
  url       = {}
}

@inproceedings{hatamizadeh2022unetr,
  author    = {Hatamizadeh, Ali and Tang, Jin and Nath, Vishwesh and Yang, Dong and Myronenko, Andriy and Landman, Bennett and Roth, Holger R. and Xu, Daguang},
  title     = {UNETR: Transformers for 3D Medical Image Segmentation},
  booktitle = {Proceedings of the IEEE/CVF Winter Conference on Applications of Computer Vision (WACV)},
  year      = {2022},
  pages     = {1743--1752},
  doi       = {10.1109/WACV51458.2022.00185},
  publisher = {IEEE},
  url       = {https://openaccess.thecvf.com/content/WACV2022/html/Hatamizadeh_UNETR_Transformers_for_3D_Medical_Image_Segmentation_WACV_2022_paper.html}
}

@inproceedings{wu2024medsegdiff,
  author    = {Wu, Junde and Luo, Lequan and Li, Yifan and Huang, Huaiqiu and Li, Xiaoyang and Wang, Yizhou and Li, Hongsheng},
  title     = {MedSegDiff: Medical Image Segmentation with Diffusion Probabilistic Model},
  booktitle = {Medical Imaging with Deep Learning (MIDL)},
  year      = {2024},
  publisher = {PMLR},
  pages     = {},
  url       = {}
}

@inproceedings{liu2020shapeaware,
  author    = {Liu, Quande and Dou, Qi and Heng, Pheng-Ann},
  title     = {Shape-aware Meta-learning for Generalizing Prostate MRI Segmentation to Unseen Domains},
  booktitle = {International Conference on Medical Image Computing and Computer-Assisted Intervention (MICCAI)},
  year      = {2020},
  publisher = {Springer International Publishing},
  address   = {Cham},
  pages     = {},
  series    = {Lecture Notes in Computer Science},
  volume    = {},
  doi       = {},
  url       = {}
}

@misc{zhang2019when,
  author        = {Zhang, Ling and Zhang, Guanhang and Wang, Pengtao and Metaxas, Dimitris N.},
  title         = {When Unseen Domain Generalization is Unnecessary? Rethinking Data Augmentation},
  year          = {2019},
  eprint        = {1906.03347},
  archivePrefix = {arXiv},
  primaryClass  = {cs.CV},
  note          = {arXiv preprint arXiv:1906.03347}
}

\end{document}